\documentclass[useAMS,usenatbib]{mn2e}
\usepackage{graphicx}
\usepackage{amssymb}
\usepackage[usenames]{color}
\title[A study on the universality and linearity of the Leavitt law in the LMC and SMC 
galaxies]{A Study on the Universality and Linearity of the Leavitt Law in the LMC and SMC 
Galaxies}

\author[A. Garc\'ia-Varela, B. E. Sabogal and M. C. Ram\'irez-Tannus]
{Alejandro Garc\'ia-Varela\thanks{E-mail: josegarc@uniandes.edu.co},  
Beatriz E. Sabogal and Mar\'ia  C. Ram\'irez-Tannus\\
Universidad de los Andes, Departamento de F\'{\i}sica,
Cra. 1E No. 18A-10, Edificio Ip, A.A. 4976, Bogot\'a, Colombia\\
}

\date{Released 2013 February 19}

\pagerange{\pageref{firstpage}--\pageref{lastpage}} \pubyear{2013}

\def\LaTeX{L\kern-.36em\raise.3ex\hbox{a}\kern-.15em
    T\kern-.1667em\lower.7ex\hbox{E}\kern-.125emX}

\begin{document}

\label{firstpage}

\maketitle

\begin{abstract}
The universality and linearity  of the Leavitt law 
are hypotheses  commonly adopted in studies of galaxy distances using 
Cepheid variables as standard candles.
In order to test these hypotheses, 
we obtain slopes  of the Leavitt law using linear regressions
of  fundamental-mode Cepheids observed by the 
Optical Gravitational Lensing Experiment project in the 
Magellanic Clouds. We find that the slopes  in $VI$-bands and in the 
Wesenheit index behave exponentially, indicating  non-linearity. 
We also find that the slopes obtained using long-period 
Cepheids can be considered as universal in the $VI$-bands, 
but not in the Wesenheit index.
\end{abstract}
\section{Introduction}
During the first decade of the twentieth century, observations of variable stars
from the Small Magellanic Cloud (SMC) galaxy taken with the 24-inch Bruce telescope at Harvard Boyden
Station in Arequipa, Peru, conducted to Henrietta Leavitt to make an important
discovery: the correlation between the logarithm of the
pulsation periods of Cepheid variables and their magnitudes  \citep{b22}.
In honour of its discoverer,  this correlation, commonly called  Period-Luminosity 
(PL) relation, has been renamed recently as the Leavitt law (LL) \citep{b23}. 
Originally, the size of the sample
used by Leavitt to establish the PL relation was
statistically small: only 25 Cepheids. Thirteen years later, the number 
of SMC Cepheids used to establish the PL relation 
increased by more than a factor of four \citep{b24}. More recently,
microlensing projects, as the Massive Compact Halo Objects (MACHO) and the
Optical Gravitational Lensing Experiment (OGLE), 
discovered hundreds of Cepheids,
as a result of monitoring systematically the bars of 
the Magellanic Clouds. Based on
these observations, it was possible to determine, with high precision, periods 
and mean \textit{VI} magnitudes of Cepheids, and the slope and zero-point values
of the LL for the Large Magellanic Cloud (LMC) galaxy \citep{b1}. These advances confirmed the 
LL as an useful modern tool to determine extragalactic distances,
tied to an assumed LMC distance modulus of $18.50$ mag \citep{f12}, 
as shown by several studies in the last
years, but in particular through the work of the Araucaria Project by measuring
distances to selected galaxies of the Local Group and the Sculptor Group
(\citealt{b19}, \citealt{a1}, and references therein).\\
In order to measure distances at galaxies through the LL, two important
hypotheses have to be adopted: its universality and linearity. Universality
means that the slope of the LL is independent 
(or weakly dependent) on metallicity, implying that this value
in a photometric band is the same for all galaxies.  
Linearity means that there is a 
linear correlation between mean magnitudes
and the logarithm of the pulsation periods of Cepheids. \\
The influence of metallicity on the PL relation has been 
discussed by several groups. Studying some selected galaxies,
\citet{u1} and \citet{b19} found strong 
evidence  in favour of the universal slope of the PL 
relation in optical bands,  between the metallicity 
range of $-1.0$ dex (IC1613) up to $-0.3$ dex (LMC).
\citet{b19} and  \citet{b12} showed that the use
of the Wesenheit index, $W_I=I-1.55(V-I)$, minimises not  only the effects of reddening  but
also those of metallicity on the PL relations. \citet{g1} and \citet{f1}
showed that the slopes of the  PL
relations in \textit{VI}-bands and in the $W_I$ index do not change
significantly in the Milky Way and LMC galaxies. \\
Although the universality and linearity of the PL relation have been assumed in
distance determination studies, there are works showing that these hypotheses
are not always truthful. With respect to the non-universality, \citet{b27},
using the infrared surface brightness technique, found that the slopes and zero
points of the PL relations of the LMC and Milky Way present a
significant dependence on metallicity  in the Wesenheit
index, a weak dependence on metallicity  in the optical \textit{VI}-bands,
and do not present any dependence on metallicity in  $K$-band. \\
With respect to non-linearity, the Russian astronomer Boris Kukarkin 
found in the thirties that the PL relations of external galaxies present 
a break around $10$ days, as it was mentioned by \citet{b25} 
and references therein. \citet{tsr01} found a break of the Period-Colour (PC)
relation at $10$ days in the LMC galaxy.  \citet{bb19} confirmed these break   
based on the statistical $F$-test. \\
From a study of the LMC Cepheids,  \citet{r5} detected  
non-linearity of the LL. This was found when using the testimator 
method, that determines if there is a slope change  in a $nth$ 
subset with respect to a smoothed slope obtained in all previous subsets.
The smoothing process is applied in order to avoid 
the outliers influence on the slopes computed. 
Applying this method, those authors detected a significant 
change in the slope  between subsets of period around $10$ days.\\
\citet{r4} presented a study of non-linearity of the  PL relation and the Period-Luminosity-Colour (PLC) 
relation using two non-parametric methods:  linear regression residuals and additive 
models. These authors found again that the PL relation is non-linear and presents a break 
around $10$ days. Possible causes for the non-linearity of 
the LL were suggested by \citet{K2004} and \citet{K2006}.\\ 
\citet{r2} studied the PL relation at individual phases of  
the pulsation cycle of Cepheids (the multi-phase approach) and  
found again a break around $10$ days.
They showed that the behaviour of the LL at mean light 
is the average of the behaviour at all phases. 
This result reinforce the use of the PL relation at mean light in the 
distance determination context. Readers interested in more details of 
these statistical methods are referred to the papers cited above. \\\\
Despite the results of these studies, universality and linearity of the LL
remain to be adopted. The slope and zero point of the LL in optical 
\textit{VI}-bands have
been established with a high degree of accuracy from  fundamental-mode Cepheids
in the LMC galaxy by \citet{b41}. When the ordinary least-squares
(OLS) regression was applied to OGLE-II data of Cepheids,  
over a range of $log P$ from  $0.4$ up to $1.5$ for the LMC, 
and from  $0.4$ up to $1.7$ for the SMC, 
Udalski realised that the standard deviation  of the residuals of 
the LMC data was almost two times smaller than that of the SMC. 
Despite that the number of Cepheids in the SMC doubles that of the LMC,
the greater dispersion of the SMC PL relation is caused by the
spatial distribution of  Cepheids in the galaxy bar,  whose thickness is placed
along the line-of-sight  with a typical depth of $\sim 0.25$ mag \citep{zz1}.
This fact, and the manner in which the PL 
relation of the LMC populates for periods greater than $2.5$ days, led  
to adopt the LMC slope value as a universal quantity \citep{b41}. \\\\
These facts and results we have mentioned above, motivate some questions behind this paper:
what is the slope of the LL from a larger sample of Cepheids? 
More specifically, what is the slope if the Cepheids in the LMC and SMC 
are considered as a single sample? \\\\
In order to test the universality and linearity hypotheses of the LL and
answer the previous questions,  this paper is structured as follows:
a description of  OGLE-II Cepheid data of the Magellanic Clouds  is presented in 
the second section.  The extension of the OLS regression, that we develop to calculate
the slope of the LL from Cepheids in the Magellanic Clouds simultaneously,
is found in the third section. Tests of universality and  linearity   of the LL,
as also our main results and a discussion of them, are presented in the fourth section.
Finally, our main conclusions are given in the fifth section.
\section{Data}
In a study of the  PL relation in optical 
bands using LMC OGLE-III fundamental-mode Cepheids, 
\citet{r1}  found a discrepancy in 
the slopes computed by them  and those reported by other authors, 
due to the different number of Cepheids used in each study. 
For this reason, to avoid effects of the sample size, 
in our study we use OGLE-II data to establish a direct  comparison between our slopes and  
those obtained by \citet{b41}, that remain as the accepted universal values. \\
The OGLE II photometric data for the Magellanic 
Clouds were obtained between 1997 and  1999, (\citealt{uu1}; \citealt{o1}). 
High quality \emph{BVI}
observations of hundreds of variable stars along the galaxies bars were
collected with the $1.3$-m Warsaw telescope, at Las Campanas Observatory, 
Chile \citep{b1}. A number of epochs between 120 to 360 was obtained in the
\emph{I}-band and between 15 to 40 in the \emph{BV}-bands. A search for
periodicity in the photometric time-series of the stars was made, using the
analysis of variance algorithm \citep{b5}. The Cepheid variables were selected
based on visual inspection of their light curves and their location on the
Colour-Magnitude diagram \citep{b2}. 
To select  fundamental-mode Cepheids, an
analysis of Fourier coefficients was performed. \\
From the  OGLE-II on-line 
archive we select  a number of 745 and 1287 fundamental-mode Cepheids,
belonging to the LMC and SMC, respectively. Their dereddened mean magnitudes 
were computed once the extinction effect was measured
(\citealt{b2}, \citealt{b3}). To compute the $A_I$ extinction coefficient,
it was determined the $I$-band magnitude of Red Clump stars in many lines-of-sight. 
Differences of the observed $I$-band magnitude of Red Clump
stars were assumed as differences in the $A_I$ extinction coefficient. To
compute the colour excess $E(B-V)$ and the extinction coefficient $A_V$  in each
line-of-sight, they adopted the reddening law of \citet{b6}.
\section{Linear Regressions}
To test the universality and linearity hypotheses of the LL 
we are interested in obtaining the slopes of the PL relations in optical bands
without adopting the values of the slopes reported by \citet{b41}. With this purpose, 
we apply two different approaches of linear fits: the OLS regression and
the multiple least-squares (MLS) regression, that is an extension 
of the OLS regression developed by us. The MLS regression
is a similar approach to the  testimator method described in the 
introduction section, but with these method 
we want to find the slope of the PL relation 
when the mathematical union of data sets of Cepheids 
from different galaxies is used, under the hypotheses 
that all data sets share the same 
slope (universality hypothesis), but each
one has a different zero point. \\
Before presenting the MLS regression, 
we briefly recall how the OLS regression works, 
emphasising on some useful results.
\subsection{Ordinary Least-Squares Regression}
\citet{b7} stated that the OLS regression should be used
to predict the value of one variable from the measurement of
another, under the following conditions:\\
(i) The nature of the linear regression scatter
is not understood, and this scatter is always greater than
the errors of the measured variables.\\
(ii) It is well established that one variable is the  cause 
and the other is the effect.\\
As a first approximation, this is the  case of the LL: we want to predict the mean 
magnitudes of Cepheids from their measured periods. To
do this, it is necessary to use the OLS regression over a set of $N$ Cepheids, each
one of them characterised by two variables: $\log{P_k}$ and
$\left<m\right>_{k}$, where the $k$ sub-index takes values between
$1$ to $N$. The logarithm of the Cepheid period is $\log{P_k}$, 
and in the context of this study, $\left<m\right>_{k}$ is the dereddened 
mean magnitude in a photometric band.
The residuals  of the measurements of dereddened mean magnitudes, 
$d_k$, are defined as the square of the differences between the 
observed values of these mean magnitudes and the
values predicted by the OLS  regression, as follows:
\begin{equation} 
d_k={(\left<m\right>_{k}- \eta \log{P}_{k} -\xi)^2} .
\label{eq1}
\end{equation}
The standard deviation
of the residuals is given by:  
\begin{equation}  
 \sigma _{\left<m\right>} = \sqrt{\frac{S}{(N-2)}}  ,
\label{eq2}
\end{equation}
where $S$ is the sum of the residuals $d_k$.
The expressions of the zero point, $\xi$, and the slope, $\eta$,
can be found by minimising $d_k$. The well know results are:
\begin{equation}
\xi=\frac{1}{N}\left(\sum_{k=1}^{N}{\left<m \right>_{k} -
\eta\sum_{k=1}^{N}{\log{P}_{k}}}\right) ,
\label{eq1b}
\end{equation}
\begin{equation}
\eta =\frac{{\sum_{k=1}^{N}{( \log{P}_{k} \left<m
\right>_{k})}}- \frac{1}{N}(\sum_{k=1}^{N}{\log{P}_{k}) 
(\sum_{k=1}^{N} \left<m\right>_{k})}}
{\sum_{k=1}^{N}{(\log{P}_{k})^2}-
{\frac{1}{N}(\sum_{k=1}^{N}{\log{P}_{k}})^2}} .
\label{eq1c}
\end{equation}
It is possible to express the 
standard deviations of the slope and zero point in terms of
the standard deviation of the residuals \citep{tt1}, as follows:
\begin{equation}
\sigma _\eta = \sigma _{\left<m
\right>}\sqrt{\frac{N}{\Delta}} ,
\label{eq3}
\end{equation}
\begin{equation}
\sigma _{\xi} = \sigma _{\left<m \right>} \sqrt{\frac{\sum_{k=1}^{N}{\left(\log{P}_k\right)^2}}{\Delta}} ,
\label{eq3a}
\end{equation}
where $\Delta$ is given by the following equation:
\begin{equation}
 \Delta=N\sum_{k=1}^{N}{\left(\log{P}_k\right)^2}-\left(\sum_{k=1}^{N}{\log{P}_k}\right)^2
\label{eq4}
\end{equation}
\subsection{Multiple Least-Squares Regression}
Let $G$ be a set of galaxies, each one having a different number $N_l$ of Cepheids. 
We want to find an analytical
expression for the slope of the PL relation resulting from the mathematical union of 
Cepheid data of all $G$ galaxies,
under the assumption that all data sets share the same slope
$\eta$. We call this value the common slope. We also want to find the analytical
expression for the zero point $\xi_l$ corresponding to each galaxy, 
under the assumption that 
each galaxy is at a different distance. For the $lth$ galaxy, 
the PL relation  can be described as:   
\begin{equation}
 \left<m \right>_{kl} = \eta \log{P}_{kl}+ \xi_l,  \hspace{0.5cm}
k=1,2,...,N_l \hspace{0.3cm} l=1,2,...,G.
\label{eq5}
\end{equation}
$\left<m \right>_{kl}$
is the predicted dereddened mean magnitude of the $kth$ Cepheid belonging to
the $lth$ galaxy, from its measured  period $P_{kl}$. 
In analogy with equation (\ref{eq1}), we define the residuals  of the measurements
of dereddened mean magnitudes for the $lth$ galaxy, as: 
\begin{equation}
D_l = {\sum_{k=1}^{N_l} \left(\left<m \right>_{kl} - \eta \log{P}_{kl} -
\xi_{l}\right)^2} .
\label{eq6}
\end{equation}
We also define the sum of the residuals for all galaxies as:
\begin{equation}
\Upsilon(\eta,\xi_l)=\sum_{l=1}^{G}{D_l} .
\label{eq7}
\end{equation}
To find the values of the common slope  $\eta$
and the  zero points $\xi_l$, $\Upsilon$ must be minimised  with respect to each
one of the $1+G$ variables ($\eta,\xi_l$):
\begin{equation}
\frac{\partial\Upsilon}{\partial\eta}=0   \hspace{0.4cm} , \hspace{0.4cm} 
\frac{\partial\Upsilon}{\xi_l}=0 .
\label{eq8}
\end{equation}
By solving these equations, the expression for the zero points $\xi_l$ of each galaxy
is obtained as:
\begin{equation}
\xi_l=\frac{1}{N_l}\left(\sum_{k=1}^{N_l}{\left<m \right>_{kl} -
\eta\sum_{k=1}^{N_l}{\log{P}_{kl}}}\right) ,
\label{eq9}
\end{equation}
which is a similar expression to the equation (\ref{eq1b}), 
given by the OLS regression, but with $\eta$
being the common slope which is independent of 
the zero points and that is given by the following equation: \\
\begin{equation}
\eta =\frac{\sum_{l=1}^{G}{\sum_{k=1}^{N_l}{( \log{P}_{kl} \left<m
\right>_{kl})}}-\omega}{\sum_{l=1}^{G}\sum_{k=1}^{N_l}{(\log{P}_{kl})^2}-\sum_{l=1
}^{G}{\frac{1}{N_l}(\sum_{k=1}^{N_l}{\log{P}_{kl}})^2}} ,
\label{eq10}
\end{equation}
where
\begin{equation}
 \omega=\sum_{l=1}^{G}{\frac{1}{N_l} \left(\sum_{k=1}^{N_l} \log{P}_{kl} \right)
\left( \sum_{k=1}^{N_l}{\left<m\right>_{kl}} \right)} .
\label{eq11}
\end{equation}
The equations (\ref{eq1b}) and (\ref{eq1c}) can be obtained of the equations 
(\ref{eq9}) and (\ref{eq10})
considering a unitary set of galaxies. In the case of the MLS regression it is possible to obtain an 
expression for the standard deviation of the residuals generalising the equation (\ref{eq2}), 
taking into account that it is necessary to include the data points of all PL relations. 
This  expression for the
standard deviation of the residual magnitudes can be written as:
\begin{equation}
\bar{\sigma} _{\left<m \right>} =\sqrt{\frac{\Upsilon}{\sum_{l=1}^{G}(N_l-2)}} .
\label{eq12}
\end{equation}
The standard deviations of
the common slope $\eta$ and the zero points $\xi_l$ can be obtained by
generalising the equations (\ref{eq3}) and (\ref{eq3a}):
\begin{equation}
\bar{\sigma} _\eta = \bar{\sigma} _{\left<m
\right>}\sqrt{\frac{\sum_{l=1}^{G}{N_l}}{\sum_{l=1}^{G}{\Delta_l}}}  ,
\label{eq13}
\end{equation}
\begin{equation}
\bar{\sigma} _{\xi} = \bar{\sigma} _{\left<m \right>} \sqrt
{\frac{\sum_{l=1}^{G} \sum_{k=1}^{N}{\left(\log{P}_{kl}\right)^2}}{\sum_{l=1}^{G}{\Delta_l}}}  ,
\label{eq13a}
\end{equation}
where $\Delta$ is given by equation (\ref{eq4}). \\
The above expressions are useful in order to test the universality and
linearity hypotheses of the PL relation. 
In order to achieve that, we apply the 
MLS regression on Cepheids of the LMC and SMC,
as a statistically sample of Cepheids.
In this case, the $l$ index runs up to $2$. We
will refer hereafter to the  data set formed by the mathematical union of the LMC
Cepheids with the SMC Cepheids as the $LMC + SMC$ set.\\
At this point, it is important summarise the 
main characteristics of 
the linear regression methods exposed.
The OLS regression allows to obtain the slope and zero point of the PL relation 
of a  single galaxy. By applying  this method to the SMC,  an
increasing of the dispersion of the PL relation caused by 
the distribution of the Cepheids along the galaxy bar is noted.
In order to obtain simultaneously the slope and zero points 
of the PL relations using the LMC and SMC  Cepheids, and damping the effect of 
the dispersion caused by the SMC Cepheids, we use the 
MLS regression over the  $LMC+SMC$ set.\\
This method allows us to test the universality of the LL in $VI$-bands making a 
study different  to the traditional ones: we want to establish if the slopes obtained
from the $LMC+SMC$ set are consistent with the slopes reported by \citet{b41}
for the LMC galaxy, despite of the  metallicity differences  of the Magellanic
Clouds.
\subsection{Testing the Multiple Least-Squares Regression}
Before applying the MLS regression to the observational data,
we perform some tests of the MLS regression over simulated data sets in order 
to find the common slope and the zero points of each set 
and verify the validity of the results of this regression method.
First, ten sets of PL relations, each one with
10000 data points,  are randomly generated. 
Each data set has a different zero point and the same slope. 
Then the MLS regression is applied, obtaining the same values
of slope and zero points used to
generate the PL relations. Then, random noise is added to the generated periods,
and the MLS regression is applied again. The slopes and zero points obtained using the
MLS regression are the ones expected. In these tests we do not 
observe trends in the residuals of the mean
magnitudes with respect to $\log P$.
Finally, we increase the number of sampling points by one order
of magnitude. As expected, we observe a corresponding decrease of
the standard deviations of slopes and zero points,
as well as in the dispersion of the residuals of the mean magnitudes.
As the MLS regression works properly with synthetic data, 
we apply it to the observational 
data sets of  Cepheids in the Magellanic Clouds.
\subsection{Removing outliers in the PL Relations}
\begin{figure}
\begin{center}
\includegraphics[width=7.5cm]{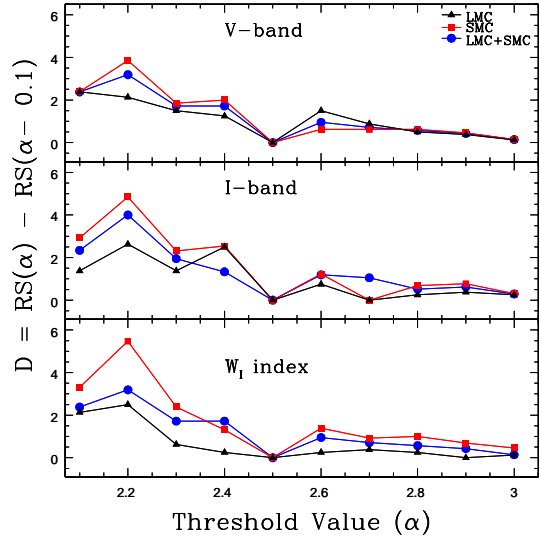}
\caption{\textbf{Percentage difference of removed stars  vs. threshold value.} 
Results for the $VI$-bands and the $W_I$ index are shown in the 
upper, middle and bottom panels, respectively. $D$ presents a local minimum in
$\alpha=2.5$. Triangles (black) correspond to the LMC Cepheid
data, squares (red) show the SMC  Cepheid data and circles (blue) correspond to 
the $LMC+SMC$ set. A colour-version of this figure is available in the 
on-line journal.}
\label{Fig1}  
\end{center}
\end{figure}
Before to obtain the slopes and zero points of the LL 
using the OLS and MLS regressions, the outliers must be discarded. 
To remove them, we find the standard deviation of the residual magnitudes, 
$\sigma _{\left<m \right>}$  and $\bar{\sigma} _{\left<m \right>}$, 
given by equations (\ref{eq2}) and (\ref{eq12}).
Then, points with a dispersion larger than $\alpha \sigma _{\left<m \right>}$ 
and $\alpha \bar{\sigma} _{\left<m \right>}$, are removed. 
In order to find the optimum threshold $\alpha$, we 
study the behaviour of  the percentage difference of 
removed stars vs. $\alpha$, selecting  $\alpha$ between
$2$ and $3$, with steps of $0.1$. 
Denoting by $RS(\alpha)$ the percentage of removed stars with threshold $\alpha$, 
and $RS(\alpha-0.1)$ the percentage of removed stars with threshold $\alpha-0.1$, 
we define the percentage difference
as: $D=RS(\alpha)-RS(\alpha-0.1)$. Fig.~\ref{Fig1} shows a plot of $D$ vs.
$\alpha$ for the LMC, SMC, and the $LMC+SMC$ set, in the \textit{VI}-bands and the 
$W_I$ index.
By examining this figure, it is clear that the local minimum in the studied
range lies in $\alpha=2.5$. For this minimum value there are 
equal numbers of  rejected points between the adjacent threshold values
of $2.4$  and $2.5$,
indicating that it is the optimum threshold value.
It is important to note that this threshold value is the same for 
the three bands studied, and is equal to the value reported by \citet{b2}.    
\subsection{Selecting Appropriate Ranges of Periods}
\begin{figure}
\begin{center}
\includegraphics[width=7.5cm]{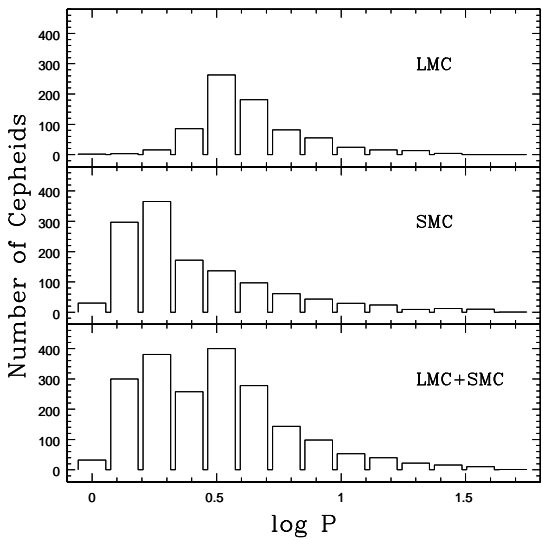}
\caption{
\textbf{Histograms of log P for Magellanic Clouds Cepheids observed by the
OGLE-II project.} The LMC sample presents a sudden decrease of detected 
Cepheids for $\log{P} \lesssim 0.4$ (upper panel).
For the SMC and the $LMC+SMC$ set, the number of detected Cepheids decreases  
for $\log{P} \lesssim 0.2$ (middle and bottom panel).
}
\label{Fig2}  
\end{center}
\end{figure}

In order to obtain slopes and zero points of the PL relations, it is necessary to establish
the range of periods that should be selected to apply the linear regressions.
Since  Cepheids with shortest periods are faintest, they may have been difficult to detect
with the 1.3-m Warsaw telescope, used by the OGLE project. Cepheids with longest periods, 
that are too bright, were over-exposed.
Fig.~\ref{Fig2} shows  a strong decrease 
in the number of detected Cepheids in the LMC galaxy about $\log{P} = 0.4$, and 
in the SMC  galaxy and the $LMC+SMC$ set about $\log{P}=0.2$.
The saturation level of the CCD used by the OGLE-II project, determined the 
upper limit of detection in $\log{P}=1.5$ for the LMC and 
$\log{P}=1.7$ for the SMC \citep{b1}.
\subsection{Discarding Cepheids with $\log P > 1.5$}
We also study Cepheids with periods larger
than the measured by the OGLE-II project ($\log P > 1.5$), 
using photometric data from the third phase of the  All Sky Automated 
Survey (ASAS-3, \citealt{p1}). This data base
contains photometry from almost ten years of continuous observations. The ASAS-3
\textit{VI} magnitudes are near to the  Johnson's  standard system, but they
were computed without include colour terms in their 
calibration.\footnote{http://www.astrouw.edu.pl/asas/explanations.html}
The number of Cepheids that we find with periods between 
$30$ and $100$ days is $19$ and $7$ for the LMC and SMC, 
respectively. Our search for periodicity is made using the analysis of
variance algorithm by \citet{b5}. \\
According to the photometric accuracy and the 
phase coverage, we try fit each light curve by Fourier 
series of orders ranging from $3$ to $6$.
The mean magnitudes  are computed following the method described by \citet{b16}. 
For each one of the ASAS-3 Cepheids,  the best fit was obtained using a Fourier series 
of $4th$ order. The mean magnitudes obtained from other orders, produce very small 
effects on the LL. We determine that the changes induced on the zero points and  
slopes by using these mean magnitudes are of the order of hundredths.\\
When the ASAS-3 Cepheids are included in the PL relations, 
they are placed too far from the trend of Cepheids observed by the OGLE-II project. 
There are two main causes that could explain this behaviour. The first one is 
the difficulty to correct by extinction, because these Cepheids are placed 
far from the galaxies  bars, where the extinction has been 
determined with less accuracy. 
The second one is the large errors (up to several tens of
magnitudes) found in the transformation of the instrumental
magnitudes to the standard system for some stars \citep{pp2}.
Based on these causes, we decide not to include 
the ASAS-3 Cepheids in our study.\\
\citet{kar} calculated the slope of the PL relation from 65 LMC Cepheids, selected 
from the ASAS catalogue of variable stars.
They found a slope value of 
$\eta _{ASAS}=-2.366 \pm 0.166$  $mag/dex$ for the $V$-band. This
value differs in more than $10\sigma$ from the value reported 
by \citet{b41} and from the value of the slope 
calculated by us with MLS regression (see Table \ref{Tab3}). This fact supports 
our decision to exclude these Cepheids of the sample.    
\section{Results and Discussion}
In order to test the universality and linearity hypotheses of the LL, 
we calculate the slopes of the PL relations for the Magellanic Clouds 
Cepheids through two different approaches. By using the OLS regression it is obtained the 
slope of the PL relation of each galaxy. By using the MLS regression we 
obtain the slope value of the LL  from Cepheids belonging to  the LMC + SMC set.
\subsection{Testing the Universality Hypothesis of the Leavitt Law}
The  universality hypothesis of the LL can be tested by showing
that the value of the slope in a photometric band is the same regardless
the metallicity of the sample of Cepheids used to derive it. 
We compute the slope of the Leavitt law for the $LMC+SMC$ set using
the MLS regression, assuming the linearity of this law in the $VI$-bands
and in the $W_I$ index, and using the ranges of periods mentioned in
subsection 3.5, 
$0.2<\log{P}<1.7$ for the SMC galaxy and the $LMC+SMC$ 
set, and $0.4<\log{P}<1.5$ for the LMC galaxy. 
Table \ref{Tab3} presents the slopes and their standard deviations obtained 
by us (MLS regression) and by \citet{b41} (OLS regression).
It is worthy of mentioning that the results 
obtained by \citet{b41} have been exactly re-obtained in this work. 
In order to establish a comparison
between our results and the reported by Udalski, the slope values 
$\eta_{MLS}$ are computed using 
dereddened $VI$ magnitudes.  The first column in Table \ref{Tab3} gives the
photometric band in which the LL is studied, the second and third
columns give the common slopes and their standard deviations obtained  
by using the equations (\ref{eq10}) and (\ref{eq13}). 
Columns fourth and fifth give
the slopes and their  standard deviations for the LMC obtained by \citet{b41},
applying the equations (\ref{eq1c}) and (\ref{eq3}).
Columns sixth and seventh give the total number of Cepheids in the LMC and SMC
galaxies that we use with the MLS regression, after rejecting outliers.
Our computed common slope $\eta_{MLS}$ of the LL is consistent 
at a level of $1.0\sigma$ and $2.2\sigma$
in $V$- and $I$-bands, respectively, with those reported by \citet{b41}.
This implies that the slope of the PL relation in the $VI$-bands is the 
same independently of the metallicity of the Magellanic Clouds. 
On the other hand, the slope $\eta_{MLS}$ in the $W_I$ index is not 
consistent with that of \citet{b41}.
These facts indicate that the LL is universal in the 
$VI$-bands but not in the $W_I$ index. The slope obtained in $I$-band 
is not as accurate as the reported by Udalski, probably 
due to blending effects reinforced by the bar geometry of the SMC galaxy.  
If Cepheids are blended with  red stars, this may change their mean $I$ 
magnitudes increasing the dispersion of the PL relation.
\subsection{Testing the Linearity Hypothesis of the Leavitt Law}
\begin{figure}
\begin{center}
\includegraphics[width=8.5cm]{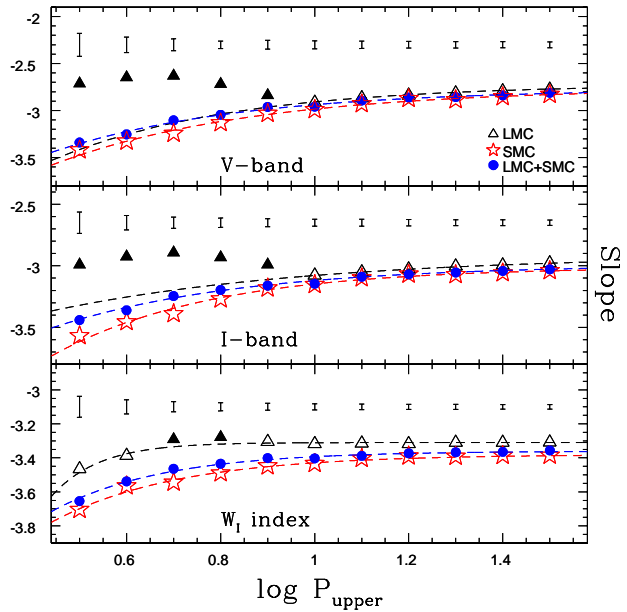}
\caption{\textbf{Slope vs. logarithm of the upper limit for range of periods.} Triangles  and stars 
represent the slopes obtained through the OLS regression 
for the LMC and SMC, respectively. 
Filled  circles  show the common slope 
obtained with the MLS regression,
for the $LMC+SMC$ set. Filled triangles  are identifying the LMC outliers data, see
text for explanation.
Dashed lines (black for the LMC, red for the SMC and blue for the $LMC+SMC$ set) 
are the exponential fits given by equation (\ref{eq17}).
Upper, middle and bottom panels show the behaviour of the slope of the LL 
for the $VI$-bands and for the $W_I$ index, respectively.
In each panel are show the slope errors 
obtained by using the MLS regression over the $LMC+SMC$ set. Large error bars corresponds 
to slopes computed with few data between the range of periods.
The slope errors corresponding to the LMC and SMC, 
are similar to the errors of the $LMC+SMC$ set. They are omitted to appreciate 
clearly the figure. A colour-version of this figure is available in the on-line journal.}
\label{Fig3}  
\end{center}
\end{figure}

The  linearity hypothesis of the LL can be tested by showing 
that the value of the slope in a photometric band is the same 
regardless of the  range of periods  selected to derive it. 
In order to compute the slope, we select the  range of periods
between a lower limit and an upper limit. 
The lower limit is useful 
to avoid the effect of Malmquist bias on the slope of the LL, 
that makes the slope value too shallow. This effect is present when the 
faintest Cepheids in a galaxy are very close to the cutoff magnitude 
of the photometry \citep{a1}.
\begin{table}
\caption{\textbf{Slopes obtained testing the universality hypothesis.} 
Columns $2$ and $4$ give the slope values obtained by us
and by \citet{b41}, respectively. 
A detailed explanation of each column is given in the text.}
\label{Tab3}
 \centering
\begin{tabular}{@{}cccccccc}
\hline
\hline
Band	&     $\eta_{MLS}$ & $\bar{\sigma}_{\eta}$    & $\eta_{OLS}$  &  $\sigma_{\eta}$ &   $N_{LMC}$ &$N_{SMC}$\\
\hline
$V$	&	-2.806  &  0.028 &	-2.775  &  0.031  & 700	&	882	\\
$I$	&	-3.024  &  0.020 &	-2.977  &  0.021  & 705	&	857	\\
$W_I$	&	-3.364  &  0.012 &	-3.300  &  0.011  & 729	&	821	\\
\hline
\end{tabular}
\end{table}
\begin{table}
\caption{\textbf{Asymptotes $\mu$ obtained testing the linearity hypothesis.} 
A complete explanation of each column is given in the text.}
\label{Tab2}
 \centering
 \begin{tabular}{@{}ccccccc}
\hline
\hline
        &\multicolumn{2}{c}{LMC}  &\multicolumn{2}{c}{SMC} &\multicolumn{2}{c}{LMC+SMC} \\
Band& $\mu$ & $\Delta\mu$   & $\mu$ & $\Delta\mu$  & $\mu$ & $\Delta\mu$ \\
\hline
$V$& -2.726 & 2.751 & -2.759 &  0.025& -2.767 & 0.062\\
$I$& -2.909 &  2.130& -3.008 &  0.029& -2.993 &  0.040\\
$W_I$& -3.311 & 0.005& -3.376 &  0.013& -3.358 &  0.009\\
\hline
 \end{tabular}
\end{table}
According to the discussion in subsection 3.5, 
we take the lower limit for the range of periods at $\log{P}=0.4$ 
for the LMC, and $\log{P}=0.2$  for the SMC and the $LMC+SMC$ set. 
We select the upper period limit of the Cepheids
ranging from $\log{P}=0.5$ up to $\log{P}=1.5$, with steps of $0.1$, 
for the LMC, SMC and the $LMC+SMC$ set.\\
Fig.~\ref{Fig3} shows the 
slope values for the PL relations as a function of the logarithm
of the upper limit of period.
Each one of these slope values for LMC and SMC 
samples was obtained by using OLS regression, and the slope 
values for the $LMC+SMC$ set were obtained using the MLS regression.
The behaviour of the obtained slopes is well fitted by an exponential function 
(dashed lines in Fig.~\ref{Fig3}):
\begin{equation}
\eta=A e^{-\beta \log{P}}+\mu  ,
\label{eq17}
\end{equation}
where the $A, \beta$, and $\mu$ parameters are obtained using a 
non-linear fitting with the IRAF task \emph{nfit1d}.
The $A$ parameter  gives information about the stretching or
squeezing of the function and the reflection about the horizontal axis. The
$\beta$ parameter gives the reflection about the vertical axis.
The asymptote $\mu$ can be interpreted as follows:
if the Cepheid population of a galaxy is observed, including 
all Cepheids with periods up to $100$ days,  and their mean magnitudes 
and periods are used in order to obtain  
the slope of the PL relation, this slope value approaches to $\mu$. \\
Inspecting the Fig.~\ref{Fig3}, it is clear that the LMC LL 
presents a break around $10$ days, in $VI$-bands. This result is 
in agreement with those reported by \citet{tsr01}, 
\citet{bb19}, \citet{r5} and \citet{r4}.\\
The LMC slopes in $VI$-bands  behave anomalously 
for short-periods ($\log P < 1.0$, black triangles in 
Fig.~\ref{Fig3}), in the sense that they move 
away from the exponential function which
adjusts the data trend for long-periods ($\log P > 1.0$). 
For this reason short-period points
for the LMC are considered to be outliers 
to fit the data by the equation (\ref{eq17}). 
On the other hand, the obtained slopes  by the OLS regression for 
the LMC are shallower than the slopes of the SMC for short-periods.
The causes of this behaviour are still under investigation by us. \\
The  behaviour of the slopes shown in 
Fig.~\ref{Fig3},  provides clear evidence of the non-linearity of 
the LL in the \textit{VI}-bands and in the $W_I$ index
for the Cepheids belonging to the Magellanic Clouds.
This result about the non-linearity of the LL is in concordance with 
the corresponding non-linearity of the PC relation, reported by 
\citet{bb19} and  \citet{r4} for the LMC Cepheids. Besides, due to 
that the PC and the PL relations are projections of the fundamental 
PLC relation \citep{r3},  our result of non-universality  of the PL 
relation suggests that the PC relation should be non-universal in the $W_I$ index.
In Table \ref{Tab2} we present the values of the asymptote $\mu$ 
and their errors obtained by using a non-linear fit with the IRAF task \emph{nfit1d}. 
The errors of the asymptote $\mu$, $\Delta\mu$, are computed taking 
into account the slope errors. 
The first column shows the photometric band where 
the LL is studied. The second and third columns give for the LMC
the values of $\mu$ and $\Delta\mu$. 
The large $\Delta\mu$ values for the LMC reflect that, 
in that case, the $\mu$ values are obtained using about one half of
the points that are used in other fits.
The following pairs of columns in Table \ref{Tab2} give the same information that the 
first two columns, but for the SMC and the $LMC+SMC$ set. \\
Despite the  non-linearity of the PL relation and its  break   around $10$ days 
(for the LMC), its use to determine distance is not significantly 
impacted by these effects due to the following reason:\\
The calculation of distance modulus of a galaxy, relative to the LMC, 
is performed from the difference between zero points of the PL relations 
of the galaxy and the LMC.
These zero points are function of the slope of the PL relation
as shown the equations (\ref{eq1b}) and (\ref{eq9}).\\
The slope of the PL relation is influenced strongly by long-period Cepheids. 
In a galaxy these Cepheids are less numerous than short-period Cepheids and 
are distributed in a wide range of periods. Therefore, a small number of them 
can produce a significant change in the 
slope value more than a large number of short-period Cepheids.\\
The effects of non-linearity 
of the PL relation can be estimated computing the deviation of the slope
from the value of $\mu$. Inspecting Fig.~\ref{Fig3}, it is seen that
the bigger is the period range used to derive the slope of the PL relation, 
the smaller is the effect of non-linearity. 
For example, in V-band the deviation of the slope obtained with MLS regression
from the value of $\mu$, in the period range from $2.5$ to $10$ days, is $3\sigma$, while for a period 
range from $2.5$ to $25$ days, is $1.1 \sigma$. \\ 
The slopes of the PL relation calculated by \citet{b41}
were obtained with Cepheids up to $32$ days (LMC) and $50$ days (SMC), for which
the effect of non-linearity  obtained in the present work is 
in the order of one sigma in V-band. 
Based on these facts our results reinforce the use of the LL to determine distances, 
independently of the non-linearity and the break of the  PL relation.\\
It can be seen from Table \ref{Tab2} that the values of the asymptotes
$\mu$ for the $LMC+SMC$ set, in $VI$-bands, agree in less than  
$1.0\sigma$ with the values for the slopes of the PL
relations found by \citet{b41} (see Table \ref{Tab3}, fourth and fifth columns).
The $\mu$ values obtained for the LMC and SMC are consistent with the values of $\eta$ 
published by \citet{b41}, up to $1.6 \sigma$ for the $V$-band, and 
$3.2 \sigma$ for the $I$-band. 
Also, it can be seen from Table \ref{Tab2} a statistically significant 
difference between the $\mu$ values of the LMC and SMC in the $W_I$ index,  
and also between the $\mu$ value of each galaxy with the $\eta$ value
published by \citet{b41}. These results suggest that the slope $\mu$  
can be considered as universal in $VI$-bands but not in 
the $W_I$ index. 
In a forthcoming paper we will be applying the OLS and MLS regressions  
in galaxies other than the Magellanic Clouds, in order to test the universality and 
linearity of the LL in these metallicity environments.
\section{Conclusions}
In this work we test the  universality 
and linearity hypotheses of the  LL, using the sample of Cepheids 
belonging to the Magellanic 
Clouds,  observed in $VI$-bands by the OGLE-II project.  In order to develop these tests, 
we compute the slope values of the LL using two different approaches. 
One of them makes a mathematical union of the Cepheid data of the LMC and SMC 
galaxies  to find a common slope applying the MLS regression. 
The other one obtains  the slope  of the PL relation  applying 
the OLS regression on a single galaxy: the LMC or the SMC. \\
The test of the universality hypothesis lead us to obtain a common slope for 
the $LMC+SMC$ set using the MLS regression. Our values are consistent with the reported by \citet{b41} 
for the LMC at a level of $1.0\sigma$ and $2.2\sigma$ in $V-$ and $I-$bands, 
respectively, and inconsistent in the $W_I$ index. Our results 
suggest a strong dependence on metallicity of the slope for the $W_I$ index, and a
weak dependence on metallicity  in the optical \textit{VI}-bands, 
in agreement with the results reported by  \citet{b27}, reaffirming the  
universality  of the LL in the $VI$-bands and showing again its 
non-universality in the $W_I$ index. \\
The test of the linearity hypothesis lead us to find that the values 
of the slopes of the LL behave exponentially as the range of periods
increases, giving clear evidence of non-linearity. In particular, we find that 
the LMC LL presents a break around $10$ days, according to
the results reported by \citet{tsr01}, \citet{bb19}, \citet{r5} and \citet{r4}.
A clear explanation of the behaviour of the LMC PL relation, in the range of 
short-periods is unknown  until now. 
Our result about the non-linearity of the LL is
in concordance with the corresponding non-linearity of the LMC PC relation, 
reported by \citet{bb19} and  \citet{r4}, and suggests that in the $W_I$ index
the PC relation should be non-universal. \\
Despite of the non-linearity of the LL, we find that the asymptote 
$\mu$ has values in $VI$-bands that are consistent with the 
slopes reported by \citet{b41} for the LMC, but inconsistent in
the $W_I$ index; therefore, in $VI$-bands, $\mu$ 
can be considered as universal and the LMC slopes given by \citet{b41} 
remain appropriate to measure extragalactic distances.
However, it is necessary to take into account that the universality of the LL 
has been tested only in a few galaxies; therefore it will be necessary to 
study the PL relation in many galaxies other than the Magellanic Clouds, 
in order to test the universality of the LL in different metallicity environments.
\section{Acknowledgments}
A. G-V and B. E. S. acknowledges support from the Facultad de Ciencias at
Universidad de los Andes through Proyecto Semilla. We would like to thank
Grzegorz Pojma\'nski, for providing Cepheids data of ASAS-3 catalogue. We also
thank our colleagues Alonso Botero and Jaime Forero, for the critical reading and important 
technical suggestions on the paper. 
Finally, we want to thank to the anonymous 
referee for the important suggestions in order to improve this work.    

\label{lastpage}
\end{document}